# CB-REFIM: A Practical Coordinated Beamforming in Multicell Networks

*Mohammad Hossein Akbari, Vahid Tabataba Vakili*

Department of Electrical Engineering, Iran University of Science and Technology (IUST), Tehran, Iran.
E-mail: mhakbari @ iust.ac.ir

**Abstract**: Performance of multicell systems is inevitably limited by interference and available resources. Although intercell interference can be mitigated by Base Station (BS) Coordination, the demand on inter-BS information exchange and computational complexity grows rapidly with the number of cells, subcarriers, and users. On the other hand, some of the existing coordination beamforming methods need computation of pseudo-inverse or generalized eigenvector of a matrix, which are practically difficult to implement in a real system.

   To handle these issues, we propose a novel linear beamforming across a set of coordinated cells only with limiting backhaul signalling. Resource allocation (i.e. precoding and power control) is formulated as an optimization problem with objective function of signal-to-interference-plus-noise ratios (SINRs) in order to maximize the instantaneous weighted sum-rate subject to power constraints.

   Although the primal problem is nonconvex and difficult to be optimally solved, an iterative algorithm is presented based on the Karush-Kuhn-Tucker (KKT) condition. To have a practical solution with low computational complexity and signalling overhead, we present CB-REFIM (coordination beamforming-reference based interference management) and show the recently proposed REFIM algorithm can be interpreted as a special case of CB-REFIM. We evaluate CB-REFIM through extensive simulation and observe that the proposed strategies achieve close-to-optimal performance.

*Index terms*: MISO-OFDMA, multicell resource allocation, power control, coordination Beamforming, interference management.

# 1 INTRODUCTION

Many researchers from networking and financial sectors have predicted that global mobile data traffic will increase 13-fold between 2012 and 2017 and will grow at compound annual growth rate (CAGR) of 66 per cent from 2012 to 2017, reaching 11.2 Exabyte per month by 2017 [1]. To fulfil this enormous mobile data traffic, the traditional cellular network that contains a restrictive number of high power base stations (BSs) has started to shift to the so-called heterogeneous network (HetNet) [2]. In HetNet, each cell includes potentially a lot of densely arranged BSs such as macro/micro/pico/femto BSs to obtain sufficient coverage for hotspot and cell edge users. On the other hand, close distance of many transmitters and receivers creates significant interference, which if not properly managed, can considerably reduce the system performance [3]. Among the different ways of handling interference, multiple inputs multiple outputs (MIMO) antenna and base station cooperation to take into accounts as the key solutions to the interference problem. Indeed, the 3GPP Long-Term Evolution-Advanced considers base station cooperation and MIMO techniques to manage inter-cell interference under the name of coordinated multipoint (CoMP) [4,5]. Coordinated transmission and reception in a multi-cell MIMO network is accomplished in two approaches: joint processing (JP) and coordinated beamforming (CB) [6]. In JP, users share data signals among the cooperating BSs in addition to control information. Inter-BS interference is suppressed by joint precoding design and simultaneously transmission among all the coordinated BSs. In this case, methods like the capacity achieving non-linear dirty-paper coding (DPC) or uncomplicated linear precoding schemes like zero-forcing (ZF) can be employed for joint transmission. However, beam vectors should be computed by centralize processing. Furthermore, this approach needs precise synchronization of BSs and exchange heavy signalling overhead on the backhaul network, especially when we have a large cooperating network. When advantage of full JP among the BSs is counteracted by the overhead, CB can be selected by the BSs as another reduced coordination scheme. In this case, the beamformers are jointly optimized among the coordinated BSs to cancel unwanted inter-BS interference. In CB, instead of user data signals, only local channel state information (CSI) and signalling information are exchanged between the coordinated BSs [3]. It is shown that multicell interference coordination can give substantial improvements in sum performance; but, joint transmission (cooperation) is very sensitive to synchronization errors and some terminals can experience performance degradations [7].

Since MIMO system benefits from receiver combining as well as transmitter beamforming, effect of interference is more destructive in the multiple input single output (MISO) than MIMO scenario. On the other hand, since mobile terminals have limitations on weight, space, processor, cost, and battery life, they are usually equipped with only one receiving antenna. Mathematically, when each mobile station has a single receiver antenna and data are not shared among base stations, the system is modelled as a MISO interference channel (IC). It has been shown that, when interference power is below a certain threshold (noisy interference), treating interference as noise is optimal [8]. In addition, even if the noisy interference condition is not satisfied, practical constraints often limit receivers to implementing single user detection (SUD) instead of genie-aided multi-user detection (MUD) [8]. Many research efforts have been focused on beam design for this MISO IC, particularly under the assumptions of linear beamforming and treating interference as noise. For example, Jorswieck *et al.* studied the basis of optimal beam vectors achieving Pareto boundary points of the rate region of the MISO IC with linear beamforming [9] and demonstrated that any Pareto-optimal beam vector at each transmitter is a combination of the matched-filtering (MF) and ZF beam vectors in the case of two users. The result was extended in [10] to general MISO interference networks using arbitrary utility functions with a monotonic property. Other noteworthy works for MISO ICs have comprised consideration of imperfect CSI [11], shared data [12], monotonic optimization with polyblock algorithm [13], second-order cone programming [14], etc. [5]. Venturino *et al.* jointly designed linear beam-vector across a set of coordinated cells and resource slots which referred as Iterative Coordinated Beam-Forming (ICBF) [15]. Although this method is an interesting iterative and near-optimal algorithm, it confronts with two difficulties: first, it needs matrix pseudo inverse computing and second it was assumed that complete instantaneous channel state information, received signal and noise power and weighted coefficients of all the users are shared among cells. On the other hand, Son *et al.* proposed an interference management (IM) algorithm, called REFIM (reference based interference management) [16] and, using the idea of *reference user*, they assumed that each BS could approximate its interference effect on users of all other cells with its interference impact on a single user to which the BS imposes the most significant interference. This idea substantially simplified the problem. However, this algorithm was proposed for single input single output (SISO) scenario without intra-cell interference and beamforming.

In this paper, based on ICBF [15], we first develop an iterative algorithm which does not need matrix pseudo inverse computing and is called iterative coordination beamforming without inverse (ICBF-WI). Then, generalization of REFIM algorithm to MISO scenario is presented which are called CB-REFIM (coordinated beamforming-REFIM). In comparison with ICBF, our algorithm has less computational complexity and lower information exchange in the cost of slight degradation of performance. Our contributions can be summarized as follows:

- Since calculating inverse of a matrix or generalized eigenvector of a matrix is difficult in practical implementation, here, we propose a simple algorithm which does not need these computations.
- We obtain a new generalization of REFIM algorithm and show REFIM can be interpreted as a special case of this algorithm. This low complexity iterative coordinated beamforming only needs a little information exchange.
- Setting lower bound of Lagrangian dual coefficients with a small positive number (e. g. $10^{-10}$) instead of zero simplifies practical implementation by neglecting effect on performance.

The remainder of this paper is organized as follows. In Section 2 we introduce the system model and the problem formulation. In Section 3 the ICBF-WI algorithm is developed and then CB-REFIM is introduced. Simulation results are given in Section 4. Finally, Section 5 concludes the paper.

*Notation*: Matrices and vectors are indicated in boldface upperspace and lowercase letters, respectively. $\mathbf{X}^H$, $\mathbf{X}^\dagger$, $tr(\mathbf{X})$ and $\|\mathbf{X}\|$ denote Hermitian, (Moore-Penrose) pseudo-inverse, trace and Frobenius norm of the matrix $\mathbf{X}$, respectively. $\mathbf{I}_n$ indicates the $n \times n$ identity matrix.

## 2 SYSTEM DESCRIPTION

In this paper, a downlink multicell MISO-OFDMA (orthogonal frequency division multiple access) wireless network composed of $M$ coordinated BSs and frequency reuse one is considered. Each base station can divide its available resources between terminals using OFDMA, which generates independent subcarriers. In addition, multiple terminals can be assigned to each subcarrier using space division multiple access (SDMA) and MISO techniques that manage co-terminal interference [7].

All BSs are equipped with $N_t$ transmit antenna and send the data simultaneously on $N$ subchannel (or orthogonal slots). We assume each subchannel is composed of a set of consecutive subcarriers which are highly correlated. It is also assumed that each single antenna mobile is served by only one BS and there is infinitely backlogged traffic in buffers, in which each BS always has sufficient data for transmission to users.

The set of users assigned to BS $m$ is denoted by $\mathcal{B}_m$. Let $s_{m,k}(n)$ be a complex symbol denoting information signal for the $k$-th user in the $m$-th cell on $n$-th subchannel and $\mathbf{v}_{m,k}(n) \in \mathbb{C}^{N_t \times 1}$ be its associated beamforming vector. It is assumed that $s_{m,k}(n)$ has independent and identical distribution (i.i.d), $E\left[|s_{m,k}(n)|^2\right] = 1$ and $E\left[s_{m_1,k_1}(n_1)s^*_{m_2,k_2}(n_2)\right] = 0$ for $(m_1,k_1,n_1) \neq (m_2,k_2,n_2)$.

The signal transmitted by base station $m$ on subchannel $n$ is as follows:

$$\mathbf{x}_m(n) = \sum_{k \in \mathcal{B}_m(n)} \mathbf{v}_{m,k}(n) s_{m,k}(n) \in \mathbb{C}^{Nt}. \tag{1}$$

The transmitted power by base station $m$ is restricted to the maximum power $P_m^{\max}$

$$\sum_{n=1}^{N} \sum_{k \in \mathcal{B}_m(n)} \|\mathbf{v}_{m,k}(n)\|^2 \leq P_m^{\max}, \quad m = 1,\ldots,M. \tag{2}$$

The discrete-time baseband received signal by user $k \in \mathcal{B}_m$ on subchannel $n$ is given by

$$y'_k(n) = \sum_{j=1}^{M} \mathbf{h}'^H_{j,k}(n) \mathbf{x}_j(n) + z'_k(n). \tag{3}$$

In (3), $\mathbf{h}'_{m,k}(n) \in \mathbb{C}^{N_t}$ represents the channel between BS $m$ and user $k$ on subchannel $n$ and includes small-scale fading, large-scale fading and path attenuation. If the coherence bandwidth of the channel is larger than the bandwidth of the each subchannel, it can be assumed that channel response at any single subchannel (slot) is flat. $z'_k(n)$ is the complex additive white Gaussian noise with variance $\sigma_k^2(n)$. Without loss of generality and to simplify the notation, we can normalize the received signal in (3) by noise standard deviation as $y_k(n) = y'_k(n)/\sigma_k(n)$, $z_k(n) = z'_k(n)/\sigma_k(n)$ and $\mathbf{h}_{m,k}(n) = \mathbf{h}'_{m,k}(n)/\sigma_k(n)$. Hence we have:

$$y_k(n) = \underbrace{\mathbf{h}_{m,k}^H(n)\mathbf{v}_{m,k}(n)s_{m,k}(n)}_{\text{desired Signal}} + \underbrace{\sum_{j=1}^{M}\sum_{\substack{u \in \mathcal{B}_j(n) \\ (j,u) \neq (m,k)}} \mathbf{h}_{j,k}^H(n)\mathbf{v}_{j,u}(n)s_{j,u}(n)}_{\text{interference}} + \underbrace{z_k(n)}_{\text{noise}}. \tag{4}$$

The channel gain matrix is defined as $\mathbf{G}_{m,k}(n) := \mathbf{h}_{m,k}(n)\mathbf{h}_{m,k}^H(n) \in \mathbb{C}^{N_t \times N_t}$. For user $k$ that receives the transmitted data from BS $m$, the signal to interference plus noise ratio (SINR) on subchannel $n$ is defined as

$$SINR_{m,k}(n) = \frac{\mathbf{v}_{m,k}^H(n)\mathbf{G}_{m,k}(n)\mathbf{v}_{m,k}(n)}{1 + \sum_{j=1}^{M}\sum_{\substack{u \in \mathcal{B}_j(n) \\ (j,u) \neq (m,k)}} \mathbf{v}_{j,u}^H(n)\mathbf{G}_{j,k}(n)\mathbf{v}_{j,u}(n)}. \tag{5}$$

According to Shannon's formula, the corresponding achievable data rate (in bits per channel use) for user $k$ in cell $m$ on subchannel $n$ is given by [17]

$$R_{m,k}(n) = \log_2\left(1 + SINR_{m,k}(n)\right). \tag{6}$$

We want to design a set of beam vectors $\mathcal{V} = \{\mathbf{v}_{m,k}(n)\}$ so as to maximize instantaneous weighted sum-rate of the system. The problem to be solved is as follows:

$$\arg\max_{\mathcal{V}} \sum_{n=1}^{N}\sum_{m=1}^{M}\sum_{k \in \mathcal{B}_m} w_k(n)\log_2\left(1 + SINR_{m,k}(n)\right), \tag{7a}$$

$$s.t. \quad \sum_{n=1}^{N}\sum_{k \in \mathcal{B}_m(n)} \|\mathbf{v}_{m,k}(n)\|^2 \leq P_m^{\max}, \quad m = 1, \ldots, M. \tag{7b}$$

Here, $w_k(n)$ is weighted coefficient which can be used to indicate different priorities depending upon fairness, different quality of service (QoS) criteria and costs. Discussing optimal policies to assign users' weight is outside the scope of this paper. If $w_k(n) = 1/MN$, the problem is simplified to the system sum-rate. Since (7) is a non-concave combinatorial optimization problem, finding a global solution through an exhaustive search has NP-hard complexity [3], which may not be tractable in practice. Therefore, in a practical situation, a suboptimal iterative solution with acceptable complexity is more preferable.

Primal optimization problem can be relaxed by transferring power constraint to the objective in the form of weighted sum. Therefore, partial Lagrangian function can be formed as

$$L(\mathcal{V},\boldsymbol{\lambda}) = \sum_{m=1}^{M}\sum_{n=1}^{N}\sum_{k\in\mathcal{B}_m(n)} w_k(n) R_{m,k}(n) \\ + \sum_{m=1}^{M} \lambda_m \left( P_m^{\max} - \sum_{n=1}^{N} \sum_{k\in\mathcal{B}_m(n)} \|\mathbf{v}_{m,k}(n)\|^2 \right), \tag{8}$$

where $\boldsymbol{\lambda} := (\lambda_1,\lambda_2,\ldots,\lambda_M)^T$ is vector of nonnegative Lagrangian multiplier. Since minimization of the so-called dual problem $\min_{\boldsymbol{\lambda}} g(\boldsymbol{\lambda}) = \min_{\boldsymbol{\lambda}} \max_{\mathcal{V}} L(\mathcal{V},\boldsymbol{\lambda})$, with respect to $\lambda_m \geq 0, m=1,\ldots,M$, is always convex regardless of the convexity or non-convexity of primal problem, the optimal solution must meet the first-order necessary conditions of optimality or Karush-Kuhn-Tucker (KKT) conditions [18]. Notice that in a maximization problem, the solution of its dual problem is an upper bound solution for that optimization problem and there is a duality gap. For convex problem this duality gap is zero and the condition is also sufficient, however this gap is not zero for a nonconvex problem. It was shown that in a multicarrier system if the number of subcarriers is large, this gap goes to zero [19]. For example in networks with a number of subcarriers equal to 64, a duality gap of less than $10^{-5}$ can be achieved, which is acceptable in practice. These KKT conditions for optimal solution of $\mathcal{V},\boldsymbol{\lambda}$ (or $\mathcal{V}^*,\boldsymbol{\lambda}^*$) briefly consist of:

i. *Feasibility of* $\mathbf{v}_{m,k}(n)$,

$$\sum_{n=1}^{N}\sum_{k\in B_m(n)} \|\mathbf{v}_{m,k}(n)\|^2 \leq P_m^{\max}, \quad m=1,\ldots,M, \tag{9a}$$

ii. *Dual feasibility of* $\lambda_m$

$$\lambda_m \geq 0, \quad \forall m=1,\ldots,M, \tag{9b}$$

iii. *Complementary slackness*

$$\lambda_m \left( P_m^{\max} - \sum_{n=1}^{N}\sum_{k\in B_m(n)} \|\mathbf{v}_{m,k}(n)\|^2 \right) = 0, \quad \forall m=1,\ldots,M, \tag{9c}$$

and finally

iv. *Optimal solution should be stationary points or*

$$\nabla_{\mathbf{v}_{m,k}(n)}\left(L(\mathcal{V},\boldsymbol{\lambda})\right) = 0, \quad \forall m=1,\ldots,M, n=1,\ldots,N, \tag{9d}$$

which $\nabla$ is defined in terms of partial derivative operator. Exerting the last condition (9d) on Equation (8) results in

$$\mathbf{T}_{m,k}(n,\lambda_m)\mathbf{v}_{m,k}(n) = \frac{w_k(n)\mathbf{G}_{m,k}(n)\mathbf{v}_{m,k}(n)}{1+\mathbf{v}_{m,k}^H(n)\mathbf{G}_{m,k}(n)\mathbf{v}_{m,k}(n)+i_{m,k}(n)}, \quad (10)$$

$$k \in B_m(n),\ m=1,\ldots,M,\ n=1,\ldots,N,$$

where $\mathbf{T}_{m,k}(n,\lambda_m) \coloneqq \mathbf{L}_{m,k}(n) + \lambda_m \ln 2 \, \mathbf{I}_{N_t}$ and

$$i_{m,k}(n) \coloneqq \sum_{l=1}^{M} \sum_{\substack{s \in \mathcal{B}_j(n) \\ (l,s) \neq (m,k)}} \mathbf{v}_{l,s}^H(n)\mathbf{G}_{l,k}(n)\mathbf{v}_{l,s}(n) \quad (11)$$

is cochannel interference received by user $k \in \mathcal{B}_m(n)$ on subchannel $n$ and

$$\mathbf{L}_{m,k}(n) \coloneqq \sum_{j=1}^{M} \sum_{\substack{u \in \mathcal{B}_j(n) \\ (j,u) \neq (m,k)}} q_{j,u}(n)\mathbf{G}_{m,u}(n) \in \mathbb{C}^{N_t \times N_t} \quad (12)$$

is defined as a leakage matrix with

$$q_{j,u}(n) \coloneqq \frac{w_u(n)\,SINR_{j,u}(n)}{1+\sum_{l=1}^{M}\sum_{s \in \mathcal{B}_j(n)} \mathbf{v}_{l,s}^H \mathbf{G}_{l,u}(n)\mathbf{v}_{l,s}(n)}, \quad u \in \mathcal{B}_j(n). \quad (13)$$

$\mathbf{L}_{m,k}(n)$ represents the amount of interference caused by BS $m$ to other cochannel users in all cells on subchannel $n$ while BS $m$ serving to user $k \in \mathcal{B}_m(n)$. Difference of leakage and interference is as follows: leakage is the signal produced by one BS which undesirably reaches all other users, while interference is defined as the unwanted transmitted power from all other BSs to one user.

## 3 THE PROPOSED ALGORITHM

Here, based on the above KKT conditions, an iterative algorithm is presented. In the first part of this section we propose an iterative algorithm. Preference of this algorithm over ICBF [15] is in lack of need for pseudo inverse, which reduces its computational complexity and simplifies its implementation. Then in the sequel of this section, CB-REFIM algorithm is proposed and its complexity and required feedback are studied. Finally, Multi-reference users' scheme is discussed and combination of CB-REFIM and fractional frequency reuse (FFR) is suggested.

A. *Coordinated beamforming without inverse*

*Proposition 1.* Near-optimal beam-vectors for nonconvex beam-forming problem (7) are of form

$$\mathbf{v}_{m,k}(n) = \frac{\beta_{m,n}(n;\lambda_m)}{\lambda_m \ln 2}\left(\mathbf{I}_{N_t} - \frac{\mathbf{L}_{m,k}(n)}{\lambda_m \ln 2 + tr(\mathbf{L}_{m,k}(n))}\right)\mathbf{h}_{m,k}(n), \quad (14)$$

where $\beta_{m,k}(n;\lambda_m) \geq 0$ and $\lambda_m > 0$ are constants to be computed iteratively. We address the obtained algorithm based on (14) as iterative coordinated beam-forming without inverse (ICBF-WI).

*Proof:* we define $\Gamma_{m,k}(n,\lambda_m) = T^{\dagger}_{m,k}(n,\lambda_m)$, where † means pseudo inverse. A solution which satisfies (10) is:

$$\mathbf{v}_{m,k}(n) = \beta_{m,n}(n;\lambda_m) \Gamma_{m,k}(n;\lambda_m) \mathbf{h}_{m,k}(n), \tag{15}$$

where $\lambda_m > 0$ and $\beta_{m,k}(n;\lambda_m)$ are constants to be computed iteratively [15]. Indeed, (10) can be rewritten as

$$\mathbf{v}_{m,k}(n) = \frac{w_k(n) r_{m,k}(n)}{1 + \|r_{m,k}(n)\|^2 + i_{m,k}(n)} \Gamma_{m,k}(n;\lambda_m) \mathbf{h}_{m,k}(n), \tag{16}$$

where $r_{m,k}(n) := \mathbf{h}^H_{m,k}(n) \mathbf{v}_{m,k}(n) = \mathbf{v}^H_{m,k}(n) \mathbf{h}_{m,k}(n)$ is a scalar. Since $r_{m,k}(n)$ itself depends on $\mathbf{v}_{m,k}(n)$, replacing (15) in (10) results in

$$w_k(n) u_{m,k}(n) = 1 + i_{m,k}(n) + (\beta_{m,k}(n;\lambda_m) u_{m,k}(n))^2, \tag{17}$$

where $u_{m,k}(n;\lambda_m) := \mathbf{h}^H_{m,k}(n) \Gamma_{m,k}(n;\lambda_m) \mathbf{h}_{m,k}(n)$. Therefore, $\beta_{m,k}(n;\lambda_m)$ can be computed by

$$\beta^2_{m,k}(n;\lambda_m) = \frac{[w_k(n) u_{m,k}(n;\lambda_m) - i_{m,k}(n) - 1]^+}{u^2_{m,k}(n;\lambda_m)}, \tag{18}$$

where $[x]^+ = \max(0,x)$. $\lambda_m$ is set so that the BS power constraint is met. By replacing (15) in constraint (7a), the following is obtained

$$\sum_{n=1}^{N} \sum_{k \in \mathcal{B}_m(n)} \|\beta_{m,n}(n;\lambda_m) \Gamma_{m,k}(n;\lambda_m) \mathbf{h}_{m,k}(n)\|^2 \leq P^{\max}_m \tag{19}$$

for $m = 1, \ldots, M$. if we define

$$f_{m,k}(n;\lambda_m) = \beta^2_{m,k}(n;\lambda_m) \|\Gamma_{m,k}(n;\lambda_m) \mathbf{h}_{m,k}(n)\|^2 \tag{20}$$

the value of $\lambda_m$ is updated so that BS power remains in feasible region

$$\sum_{n=1}^{N} \sum_{k \in \mathcal{B}_m(n)} f_{m,k}(n,\lambda_m) \leq P^{\max}_m. \tag{21}$$

Obtaining $\Gamma_{m,k}(n;\lambda_m)$ needs computing inverse of a matrix, which increases computational complexity of the algorithm. Inverse computation is a common problem in a large variety of beam forming schemes, like zero forcing family. Here we propose a new beamforming scheme which does not need taking inverse based on Sherman–Morrison formula. The Sherman–Morrison formula is a special case of the Woodbury formula [20]. Suppose **A** is an invertible square matrix and **u** and **v** are vectors. The matrix inversion lemma states that

$$\left(\mathbf{A}+\mathbf{u}\mathbf{v}^{H}\right)^{-1} = \mathbf{A}^{-1} - \frac{\mathbf{A}^{-1}\mathbf{u}\mathbf{v}^{H}\mathbf{A}^{-1}}{1+\mathbf{v}^{H}\mathbf{A}^{-1}\mathbf{u}}. \tag{22}$$

This identity is widely used in signal processing and control because it allows us to invert a rank-one extension of a non-singular matrix without requiring matrix inversion. Therefore,

$$\boldsymbol{\Gamma}_{m,k}(n;\lambda_m) = \frac{1}{\lambda_m \ln 2}\left(\mathbf{I}_{Nt} - \frac{\mathbf{L}_{m,k}(n)}{\lambda_m \ln 2 + tr(\mathbf{L}_{m,k}(n))}\right). \tag{23}$$

The proof of proposition is completed by placing (23) in (15). □

As a special case, if $N_t = 1$, Relation (23) becomes

$$\Gamma_{m,k}(n;\lambda_m) = \frac{1}{\lambda_m \ln 2 + L_{m,k}(n)}, \tag{24}$$

which is equivalent to the price or taxation term in the improved iterative water-filling (I-IWF) algorithm [21,22]. Also, if there is no coordination, the algorithm is reduced to conventional water-filling [19].

Here, we can assume following assumption: In the case of $N_t \geq MK$, $\{\mathbf{h}_{m,k}(n), \forall k\}$ are linearly independent for each $m$. Also, in the case of $N_t < MK$, element vectors of any subset of $\{\mathbf{h}_{m,k}(n), \forall k\}$ with cardinality $N_t$ are linearly independent for each $m$. This assumption is almost surely satisfied for randomly realization channel vectors [5]. It has been demonstrated in previous works [5,9] that, in MISO scenario, if $N_t \geq MK$, all available BS power is allocated and power constrains (7b) are satisfied with equality. Therefore, from complementary slackness (9c), value of $\lambda_m$ should be positive. As a practical point, in simulation and real implementation, the lower bound of $\lambda_m$ is not usually set equal to zero; instead it is set equal to a very small positive value, e.g. $\lambda_m^{\min} = 10^{-10}$. Effect of this engineering approximation is negligible, because

$$\lambda_m^{\min}\left(P_m^{\max} - \sum_{n=1}^{N}\sum_{k \in B_m(n)}\|\mathbf{v}_{m,k}(n)\|^2\right) \leq \lambda_m^{\min}\left(P_m^{\max}\right) = 10^{-10} \approx 0. \tag{25}$$

After computing $\lambda_m$ and $\beta_{m,k}(n;\lambda_m)$, beam vectors are updated by (14) and this process is iterated until convergence. Suitable value of $\lambda_m$ can be efficiency selected by bisection between $\lambda_m^{\min}$ and an upper bound $\max_n \max_{k \in \mathbf{B}_m(n)}\left\{\frac{w_k(n)}{\ln 2}\|\mathbf{h}_{m,k}(n)\|^2\right\}$ because $\|\mathbf{v}_{m,k}(n)\|^2$ is non-increasing function of $\lambda_m > 0$ [15]. Iterative algorithm of ICBF-WI is described in Table 1. Line (4) of this algorithm distinguishes it from CB-REFIM algorithm that will be discussed in the next part. This algorithm needs an initial feasible set of beam-vectors. Although a better initialization may help speed up convergence, ICBF-WI usually achieves a similar sum-rate at convergence for different initializations. Three simple initial beam vectors are:

**TABLE 1.** *ICBF-WI and CB-REFIM algorithm*

| | |
|---|---|
| 1: | Initialize $L_{in,max}$, $L_{out,max}$, $\{\mathbf{v}_{m,k}(n), k \in \mathcal{B}_m(n), m=1,\ldots,M, n=1,\ldots,N\}$; |
| 2: | $l_{out}=0$; |
| 3: | **repeat** |
| 4: |   Compute $\{\mathbf{L}_{m,k}(n), k \in \mathcal{B}_m(n), m=1,\ldots,M, n=1,\ldots,N\}$ from (12) for ICBF-WI and from (33) for CB-REFIM ; |
| 5: |   $l_{in}=0$; |
| 6: |   **repeat** |
| 7: |     compute $\{i_{m,k}(n), k \in \mathcal{B}_m(n), m=1,\ldots,M, n=1,\ldots,N\}$ according to (11); |
| 8: |     **for** *m=1* to *M* **do** |
| 9: |       compute $\lambda_m$ and $\{\beta_{m,k}(n;\lambda_m), k \in \mathcal{B}_m(n), n=1,\ldots,N\}$ from (18); |
| 10: |       Update $\{\mathbf{v}_{m,k}(n), k \in \mathcal{B}_m(n), n=1,\ldots,N\}$ by (14); |
| 11: |     **end for** |
| 12: |     $l_{in}=l_{in}+1$; |
| 13: |   **until** convergence or $l_{in}=L_{in,max}$; |
| 14: |   $l_{out}=l_{out}+1$; |
| 15: | **until** convergence or $l_{out}=L_{out,max}$; |

1- *Channel-matched (CM)* or *maximum ratio transmission (MRT) beamforming*:

$$\mathbf{v}_{m,k}(n) = \sqrt{\frac{P_m^{\max}}{NK}} \frac{\mathbf{h}_{m,k}(n)}{\|\mathbf{h}_{m,k}(n)\|} . \tag{26}$$

2- *Per-cell zero-forcing (ZF)*: Intra-cell interference can be suppressed if $N_t \geq MK$ and initial set of beam vectors is selected so as [9]

$$\mathbf{v}_{m,k}(n) = \sqrt{\frac{P_m^{\max}}{NK}} \frac{\Pi^{\perp}_{\tilde{\mathbf{H}}_{m,k}(n)}\mathbf{h}_{m,k}(n)}{\left\|\Pi^{\perp}_{\tilde{\mathbf{H}}_{m,k}(n)}\mathbf{h}_{m,k}(n)\right\|}, \tag{27}$$

wherein $\Pi_\mathbf{X} := \mathbf{X}(\mathbf{X}^H\mathbf{X})^{-1}\mathbf{X}^H$ is orthogonal projection onto the column space on $\mathbf{X}$, $\Pi_\mathbf{X}^{\perp} := \mathbf{I} - \Pi_\mathbf{X}$ is orthogonal complement of column space of $\mathbf{X}$ and $\tilde{\mathbf{H}}_{m,k}(n) := [\mathbf{h}_{m,1}(n),\ldots,\mathbf{h}_{m,k-1}(n),\mathbf{h}_{m,k+1}(n),\ldots,\mathbf{h}_{m,K}(n)]$.

3- *Maximum signal to leakage plus noise ratio (MSLNR) beamforming*: For any user $k \in B_m(n)$, SLNR on slot *n* is defined as [23]:

$$SLNR_{m,k}(n) = \frac{\mathbf{v}_{m,k}^H(n)\mathbf{G}_{m,k}(n)\mathbf{v}_{m,k}(n)}{\sum_{j=1}^{M}\sum_{\substack{u \in B_j(n) \\ (j,u) \neq (m,k)}} \mathbf{v}_{m,k}^H(n)\mathbf{G}_{m,u}(n)\mathbf{v}_{m,k}(n) + 1}. \tag{28}$$

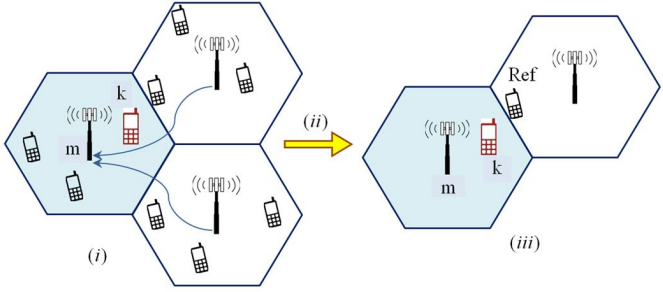

**Fig. 1** *Reference user selection method in CB-REFIM: (i) BS m receives information of users from neighboring BSs. (ii) BS m selects a reference user using (28) and (iii) BS m determines beam vector by CB-REFIM algorithm.*

It is assumed that the available power is equally split across beams in initialization; i.e. $\|\mathbf{v}_{m,k}(n)\|^2 = P_m^{\max}/NK$; then, the following maximum $SLNR_{m,k}(n)$ problem can be given:

$$\mathbf{v}_{m,k}(n) = \arg\max_{\mathbf{x}\in\mathbb{C}^{Nt}} \frac{\mathbf{x}_{m,k}^H(n)\mathbf{G}_{m,k}(n)\mathbf{x}_{m,k}(n)}{\mathbf{x}_{m,k}^H(n)\mathbf{D}_{m,k}(n)\mathbf{x}_{m,k}(n)} \quad s.t. \|\mathbf{x}\|^2 = \frac{P_m^{\max}}{NK}, \tag{29}$$

where

$$\mathbf{D}_{m,k}(n) := \sum_{j=1}^{M} \sum_{\substack{u \in B_j(n) \\ (j,u) \neq (m,k)}} \mathbf{G}_{m,u}(n) + \frac{NK}{P_m^{\max}} \mathbf{I}_{Nt}. \tag{30}$$

The solution is given in terms of eigenvector corresponding to the largest eigenvalue of the matrix $\mathbf{D}_{m,k}^{-1}(n)\mathbf{G}_{m,k}(n)$ as [23]

$$\mathbf{v}_{m,k}(n) \propto \max\ eigenvector\left(\mathbf{D}_{m,k}^{-1}(n)\mathbf{G}_{m,k}(n)\right). \tag{31}$$

Norm of $\mathbf{v}_{m,k}(n)$ is adjusted to 1[15].

*Remark 1.* Since $q_{j,u}(n)\mathbf{G}_{m,u}(n)\mathbf{h}_{m,k}(n) = c'_{j,u}(n)\mathbf{h}_{m,u}(n)$ in which $c'_{j,u}(n)$ is a complex scalar, then (14) can be rewritten as $\mathbf{v}_{m,k}(n) = \sum_{j=1}^{M}\sum_{u\in B_j} c_{j,u}(n)\mathbf{h}_{m,u}(n)$ for some scalar $c_{j,u}(n)$. This is similar to recent results has been presented for relaxed zero forcing coordinated beamforming (RZFCB) where summation is taken over leakage constraints by equality [5, lemma 3].

*B. CB-REFIM*

Although ICBF-WI is an iterative algorithm, its implementation in a real system still is a challenging problem; sharing channel gains and noise power across the cells imposes overhead to backhaul. To tackle this complexity and develop a simply distributed scheme with small feedback exchange, we use concept of *Reference user* with some modifications which was first proposed in [16]. Let $\mathbf{N}(m)$ be set of BS *m* and its neighbouring (adjacent) BSs. Also, the set of all (scheduled) users on subchannel *n* in $\mathbf{N}(m)$ is denoted by $A(m,n)$. When BS $m$ wants to compute its beam vector for user $k$ on subchannel $n$, the reference user is defined as the user among $A(m,n)$ whose channel from BS $m$ has the most inner product with channel between BS $m$ and user $k$ on subchannel $n$. From the BS $m$ point of view, although its transmit power interferes with all other users on the corresponding subchannel in all cells, the effect will be critical, especially for the user whose channel has biggest scalar projection on channel of user $k$. In other words, the reference user is the user which approximately receives the highest leakage of the transmitted signal from BS $m$ to user $k$. Therefore, we propose an online reference user selection method as follows:

$$\text{ref}_{m,k}(n) = \arg \max_{\substack{u \in A(m,n) \\ u \neq k}} \|\mathbf{h}_{m,u}(n)\|^2 \|\mathbf{h}_{m,u}^H(n)\mathbf{h}_{m,k}(n)\|^2. \tag{32}$$

Obtaining the above relation is straightforward and indeed it is the largest amount of $\|\mathbf{G}_{m,u}(n)\mathbf{h}_{m,k}(n)\|^2$. Similar to REFIM, here the term of $\mathbf{G}_{m,u}(n)\mathbf{h}_{m,k}(n)$ is the dominant term of leakage in $\mathbf{v}_{m,k}(n)$. (See (14) and (12)). This relation is somehow different from definition of reference user in REFIM algorithm [16], in which the reference user has been defined as the user within $A(m,n)$ with the strongest channel gain between the user and BS $m$, i.e. $\text{ref}_{m,k}(n) = \arg \max_{u \in A(m,n)} \|h_{m,u}(n)\|^2$. These differences return to the fact that REFIM has been proposed for a SISO scenario where scalar $h_{m,k}(n)$ is a common factor with no effect on maximization. Hence, REFIM can be interpreted as a special case of CB-REFIM in SISO system. Also, in CB-REFIM, BS $m$ is included in $\mathbf{N}(m)$ as well as neighbouring BSs because of the existence of intra-cell in addition to inter-cell interference. Fig. 1 depicts an example of reference user selection procedure.

Therefore leakage matrix in CB-REFIM can be rewritten as

$$\mathbf{L}_{m,k}(n) = q_{j,\text{ref}_{m,k}(n)}(n)\mathbf{G}_{m,\text{ref}_{m,k}(n)}(n) \in \mathbb{C}^{N_t \times N_t}. \tag{33}$$

Other relations remain the same as previous section and their proving is not needed, because the system can be modelled either as a two cell system with only one user in each cell or as a single cell system with two users when intra-cell interference is dominating. As illustrated in Table 1, CB-REFIM is similar to ICBF-WI algorithm, except that Relation (33) is used instead of (12) in Line 4 of the iterative algorithm.

*C. Complexity and feedback deduction*

Since ICBF algorithm [15] requires computing the pseudo-inverse of matrices $\{\mathbf{T}_{m,k}(n;\lambda_m)\}$, one iteration of the algorithm has complexity that is scales of $MNKN_t^3$. Since ICBF-WI algorithm does not require pseudo-inverse, it has the complexity which is scales of $MNKN_t$. Complexity of CB-REFIM is also $MNKN_t$ with a lower scale because two summations in (12) are removed in computation of CB-REFIM leakage matrix (33).

The information feedback exchange between BSs is also reduced in CB-REFIM. The ICBF needs four types of information:

(F1) $\mathbf{h}_{m,k}(n)$: Channel vector of each users in $\mathbf{N}(m)$

(F2) $w_k(n)$: Weight of each user in $\mathbf{N}(m)$

(F3) $\left\|\mathbf{v}_{j,u}^H(n)\mathbf{h}_{j,u}(n)\right\|^2$: Received signal strength of each user in $\mathbf{N}(m)$

(F4) $\sum_{l=1}^{M}\sum_{s\in B_l(n)}\mathbf{v}_{l,s}^H\mathbf{G}_{l,u}(n)\mathbf{v}_{l,s}(n)+\sigma^2$: Received noise plus interference strength of each user in $\mathbf{N}(m)$

However, in CB-REFIM, except (F1), other information (F2)-(F4) is only received from the reference user and information of other users is not required, which significantly reduce the amount of feedback.

*D. Multiple reference users*

When the number of users in each cell and number of BS antennas are increased, computational complexity of ICBF-WI is increased with respect to number of users and number of transmitter antenna. On the other hand, intra-cell and inter-cell interferences cause decreased weighted sum rate of CB-REFIM in comparison with ICBF. In this situation, considering the second reference user makes a compromise between complexity and throughput.

*E. CB-REFIM+FFR*

Since the proposed system is a multicarrier system (OFDMA: orthogonal frequency division multiple access), there is an additional degree of freedom in frequency. An interesting practical scenario in a heterogeneous network is fractional frequency reuse (FFR) [24]. Combination of CB-REFIM and FFR can also be used as a special case of CB-REFIM algorithm.

# 4 NUMERICAL RESULTS

In this section, performance of CB-REFIM is studied using Monte-Carlo simulations; since the present results are intended to be compared with those of ICBF algorithm, we have selected simulation parameters the same as [15]. A wireless network with hexagonal cells is considered in which central cluster of *M*=3 neighbouring base stations is coordinated. Interference of out-of-cluster, i.e., interference from the cells no. 4 to 27 in the first and second tiers is treated as noise. The distance between adjacent base stations is set to 2 Km. *K* mobiles are uniformly distributed around each serving BS within circular annulus of internal and external radii of 500 and 1100 meters, respectively. The baseband channel from *m*-th BS to *k*-th mobile user on subchannel *n* contains path loss, shadowing and small scale fading, is modelled as

$$\tilde{h}_{m,k}(n) = \left\{\left(\frac{200}{d_{m,k}}\right)^{3.5}L_{m,k}\right\}^{\frac{1}{2}}\bar{h}_{m,k}(n), \tag{34}$$

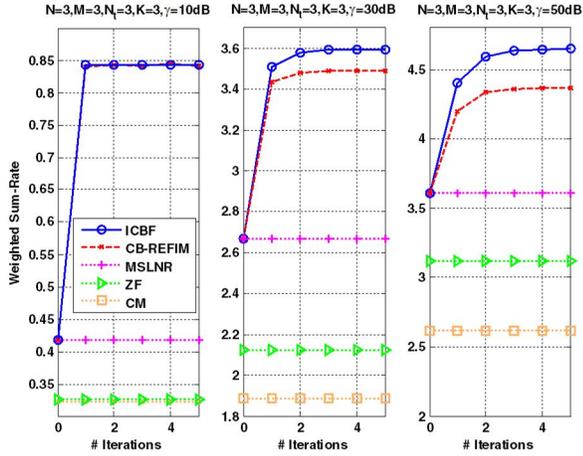 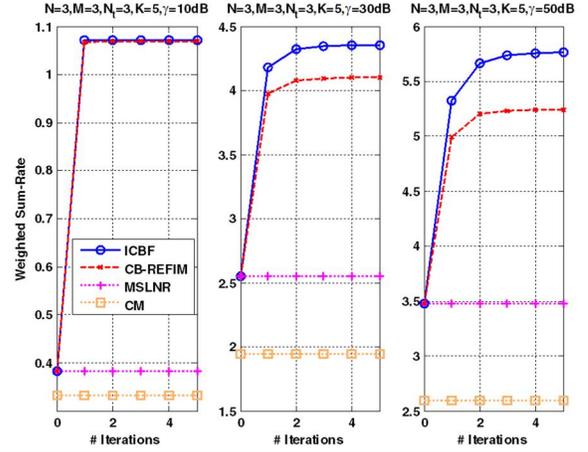

**Fig. 2** *Sum-rate of CB-REFIM algorithm versus the number of iteration. Initialization point is MSLNR. ICBF, CM, ZF and MSLNR are shown for comparison. Simulation parameters include: M=3, N=3, K=3, γ=10, 30, 50dB.*

**Fig. 3** *Sum-rate of CB-REFIM algorithm versus the number of iteration. Initialization point is MSLNR. ICBF, CM and MSLNR are shown for comparison. Simulation parameters include: M=3, N=3, K=5, γ=10, 30, 50dB.*

where $d_{m,k}$ is distance of user $k$ from base station $m$; $10\log_{10}(L_{m,k}(n))$ is a real Gaussian random variable with mean of zero and standard deviation of 8 accounting for large-scale log-normal shadowing; finally, $\bar{h}_{m,k}(n) \sim \mathbf{CN}(0,\mathbf{I}_{Nt})$ is a circularly symmetric complex Gaussian random vector accounting for Rayleigh small scale fading. Maximum total BS power is assumed equal for all BSs and all weight coefficients are selected equal ($w_k(n)=1/MN$). Noise power $N_k(n)$ at each mobile is modelled as [15]

$$N_k(n) = \sigma^2 + \sum_{m=4}^{27}\left(\frac{200}{d_{m,k}}\right)^{3.5} L_{m,k} \frac{P_{\max}}{N}, \qquad (35)$$

which includes long-term interference of uncoordinated BSs. Here, $\sigma^2$ is thermal noise power which is assumed to be the same for all receivers. Transmit signal to noise ratio is defined as $\gamma := P_{\max}/\sigma^2$. All simulations are conducted using a QPSK transmit constellation and each plot is obtained by averaging sum-rate over 100 independent random locations of users. Finally like [15], we set $L_{in,\max}=40$ and $L_{out,\max}=4$.

For estimation of channel gain, it has been assumed that the system uses cell-specific orthogonal reference signals. Users' terminal know the reference (pilot) signals of neighbouring first-tier and they can separately determine interference or channel gain [25]. These reference signals have been designed cleverly and help to decentralize the implementation of system and signalling overhead reduction.

First, a comparison is made between ICBF and CB-REFIM with other beamforming algorithms versus different values of iteration and SNRs. In Figs. 2 and 3, weighted sum-rate is illustrated versus the number of iterations of outer loop for $\gamma$=10, 30,

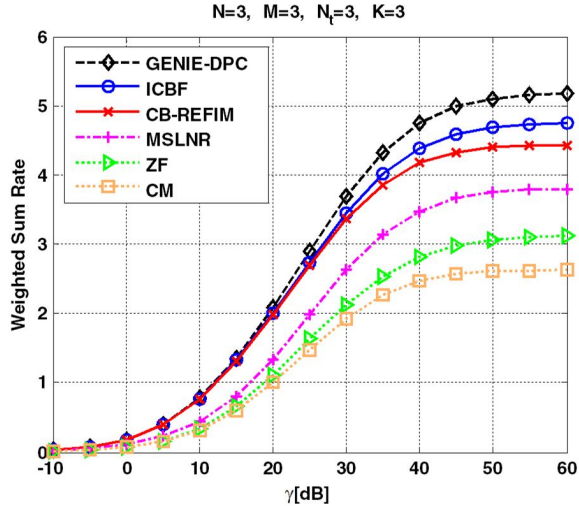 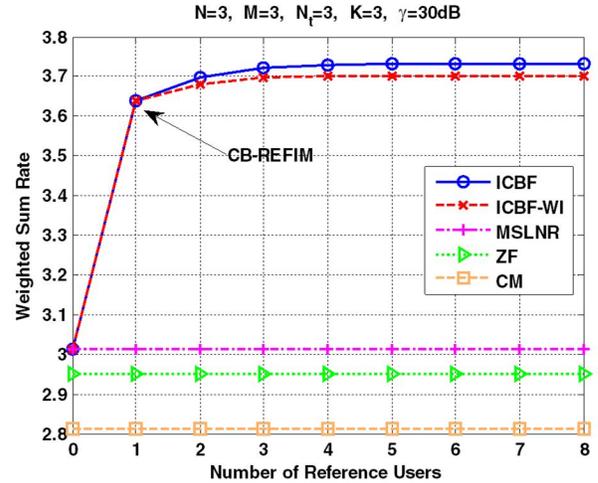

**Fig. 4** *Sum-rate versus γ. CB-REFIM algorithm is compared by ICBF, MSLNR, ZF and CM. Sum-rate is provided by in-cell DPC in the ideal case where a genie perfectly removes the co-channel interference from other coordinated cells. System parameters include N=3, M=3, $N_t$=3 and K=3 users in each cell.*

**Fig. 5** *Effect of the number of reference user on weighted sum-rate. 98% of sum-rate is achieved by considering only one reference user. Simulation parameters are M=3, N=3, K=3, γ=30 dB.*

50 dB and for the number of user in each cell $K$=3, 5, respectively. These examples indicate that only very few iterations (3-5) are sufficient for convergence. Also, although better initialization might help speed up convergence, CB-REFIM achieves a similar sum-rate at convergence for different initializations, only one of which has been drawn for better observation. As can be observed, sum-rate of CB-REFIM is slightly smaller than that of ICBF. This degradation is a low cost against significant amount of reduction in complexity and signalling overhead. However, performance is much better than that of CM, per-cell ZF and MSLNR.

In Fig. 4, sum-rate performance of the CB-REFIM algorithm at convergence versus signal to noise ratio (SNR) γ for $K$=3 is illustrated. In this plot, sum-rate provided by in-cell dirty paper coding (DPC) [26] is also reported as an ideal case for the sake of comparison, where a genie perfectly remove cochannel interference from all the coordinated cells. Indeed genie idealizes message knowledge of the existing transmitters' messages at transmitter and is used for the aim of fully interference cancellation. Notice that, this sum-rate is not achievable when coordinated base stations only share channel quality measurement. At any value of γ, CB-REFIM solution significantly improves upon the initial starting point (CM, ZF and MSLNR) and provides large throughput gains by balancing power and optimizing beam direction. For example, for $K$=3, the sum-rate provided by ICBF is roughly 42%, 31% and 24% higher than the sum-rate provided by MSLNR beamforming at γ=10, 30 and 50 dB, respectively. This improvement for CB-REFIM is 41%, 28% and 16%, respectively. Also, it is surprising that the sum-rate performance of CB-REFIM is very close to genie-DPC over a wide range of γ's and is negligible for $γ ≤ 25$ dB.

Fig. 5 shows weighted sum-rate performance of ICBF and ICBF-WI versus different numbers of reference users per subchannel and that of CM, per-cell ZF and MSLNR, as a baseline. Without a reference user, ICBF-WI is reduced as selfishly single-cell beam vector allocation. With considering reference users, higher performance gain is obtained. It is interesting that considering only one reference user per subchannel is efficient enough because it can obtain more than 90% of performance considering all $MK$-1 leaked users.

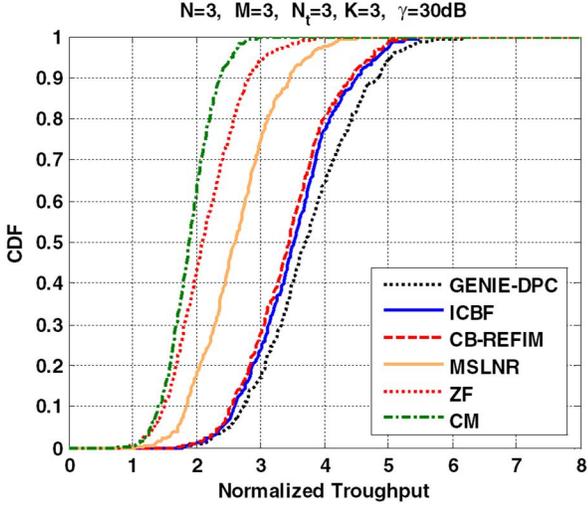 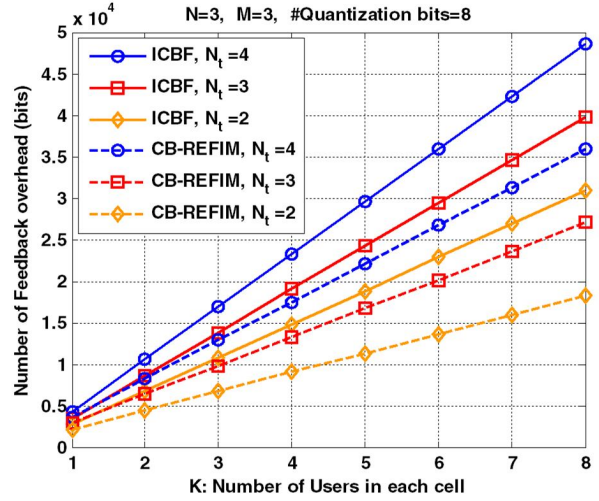

**Fig. 6** *CDF of different algorithms. Simulation parameters are M=3, N=3, K=3, γ=30 dB.*

**Fig. 7** *The number of feedback bits versus number of users in each cell for different numbers of transmitter antennas. Simulation parameters are M=3, N=3, number of quantization bits=8.*

Fig. 6 demonstrates cumulative distribution function (CDF) of individual terminals rates in the network for different schemes. Compared to CM, ZF and MSLNR, CB-REFIM can improve throughput for all users in the network. Specially, higher improvement (62% improvement with respect to CM) is observed for users achieving low throughputs, like users at cell edges. This is because the cell edge users have a determinant role on the CDF curve.

In Fig. 7, the number of required signalling bits for information exchange between BSs is depicted versus the number of users in each cell for different numbers of transmitter antennas. This feedback overhead is desired to be small as possible. Each real value was assumed to be quantized by 8 bits. As can be seen, CB-REFIM only needs 60% to 75% of feedback bits required for ICBF. Here we exchange both channel magnitude and direction information. If only channel direction information is exchanged, this overhead can be further reduced.

## 5 CONCLUSION

In this paper, linear precoding design and power control were considered for downlink multicell MISO-OFDMA interference channel with the uniform frequency reuse framework. In particular, we investigated nonconvex weighted sum-rate maximization (WSRM) of coordinated cells subject to per-cell power constraint. Since computation of pseudo-inverse or generalized eigenvector of a matrix are difficult for practical implementation in a real system, in contrast to some of the available methods for coordination beamforming which need computation of pseudo-inverse or generalized eigenvector of a matrix, we proposed a coordination beamforming scheme which needed neither computation of inverse nor generalized eigenvector.

Although primal problem was nonconvex and difficult to be optimally solved, an iterative algorithm was presented based on the Karush-Kuhn-Tucker (KKT) first order necessary conditions of optimality. These optimality properties were used to propose low-complexity strategies: both a centralized scheme and a distributed version that only required local channel knowledge and processing.

Since a practical solution with low computational complexity and signalling overhead –which are crucial for low-cost small-cell solutions– was desired, we presented coordination beamforming-reference based interference management (CB-REFIM) and the recently proposed REFIM algorithm was demonstrated to be interpreted as a special case of CB-REFIM. We evaluated CB-REFIM using extensive simulation under various configurations and it was found that, compared to the scheme without interference management (IM), CB-REFIM can achieve close-to-optimal performance but with significant reduction in complexity and signalling overhead. One of our future works will consider the robust coordinated beamforming with imperfect CSI in multicell systems.

# 6 APPENDIX

Proof of relation (10):

$$\frac{\partial L(\mathcal{V},\lambda)}{\partial \mathbf{v}_{m,k}(n)} = \sum_{j=1}^{M} \sum_{u \in B_j(n)} \frac{w_u(n)}{(1+SINR_{j,u}(n))\ln 2} \frac{\partial SINR_{j,u}(n)}{\partial \mathbf{v}_{m,k}(n)} - 2\lambda_m \mathbf{v}_{m,k}(n) = 0,$$

$$\frac{\partial SINR_{m,k}(n)}{\partial \mathbf{v}_{m,k}(n)} = \frac{2\mathbf{G}_{m,k}(n)\mathbf{v}_{m,k}(n)}{1+\sum_{j=1}^{M}\sum_{\substack{u \in B_j(n) \\ (j,u) \neq (m,k)}} \mathbf{v}_{j,u}^H(n)\mathbf{G}_{j,k}(n)\mathbf{v}_{j,u}(n)},$$

$$\frac{\partial SINR_{j,u}(n)}{\partial \mathbf{v}_{m,k}(n)} = \frac{-2\mathbf{G}_{m,u}(n)\mathbf{v}_{m,k}(n)\left\{\mathbf{v}_{j,u}^H(n)\mathbf{G}_{j,u}(n)\mathbf{v}_{j,u}(n)\right\}}{\left(1+\sum_{l=1}^{M}\sum_{\substack{s \in B_l(n) \\ (l,s) \neq (j,u)}} \mathbf{v}_{l,s}^H(n)\mathbf{G}_{l,u}(n)\mathbf{v}_{l,s}(n)\right)^2},$$

$$\frac{\partial L(\mathcal{V},\lambda)}{\partial \mathbf{v}_{k,m}(n)} = \frac{w_k(n)}{(1+SINR_{m,k}(n))\ln 2} \frac{2\mathbf{G}_{m,k}(n)\mathbf{v}_{m,k}(n)}{1+\sum_{j=1}^{M}\sum_{\substack{u \in B_j(n) \\ (j,u) \neq (m,k)}} \mathbf{v}_{j,u}^H(n)\mathbf{G}_{j,k}(n)\mathbf{v}_{j,u}(n)} - \cdots$$

$$\sum_{j=1}^{M}\sum_{\substack{u \in B_j(n) \\ (j,u) \neq (m,k)}} \frac{w_u(n)}{(1+SINR_{j,u}(n))\ln 2} \frac{2\mathbf{G}_{m,u}(n)\mathbf{v}_{m,k}(n)\left\{\mathbf{v}_{j,u}^H(n)\mathbf{G}_{j,u}(n)\mathbf{v}_{j,u}(n)\right\}}{\left(1+\sum_{l=1}^{M}\sum_{\substack{s \in B_l(n) \\ (l,s) \neq (j,u)}} \mathbf{v}_{l,s}^H(n)\mathbf{G}_{l,u}(n)\mathbf{v}_{l,s}(n)\right)^2} - \cdots$$

$$2\lambda_m \mathbf{v}_{k,m}(n) = 0,$$

$$\left\{\sum_{j=1}^{M}\sum_{\substack{u\in B_j(n)\\(j,u)\neq(m,k)}}\frac{w_u(n)\mathbf{v}_{j,u}^H(n)\mathbf{G}_{j,u}(n)\mathbf{v}_{j,u}(n)\mathbf{G}_{m,u}(n)}{\left(1+\sum_{l=1}^{M}\sum_{\substack{s\in B_l(n)\\(l,s)\neq(j,u)}}\mathbf{v}_{l,s}^H(n)\mathbf{G}_{l,u}(n)\mathbf{v}_{l,s}(n)\right)\left(1+\sum_{l=1}^{M}\sum_{s\in B_l(n)}\mathbf{v}_{l,s}^H(n)\mathbf{G}_{l,u}(n)\mathbf{v}_{l,s}(n)\right)}+\lambda_m\ln 2\mathbf{I}_{N_t}\right\}\mathbf{v}_{m,k}(n)=\cdots$$

$$\frac{w_k(n)\mathbf{G}_{m,k}(n)\mathbf{v}_{m,k}(n)}{1+\mathbf{v}_{m,k}^H(n)\mathbf{G}_{m,k}(n)\mathbf{v}_{m,k}(n)+i_{m,k}(n)},$$

$$\left(\mathbf{L}_{m,k}(n)+\lambda_m\ln 2\mathbf{I}_{N_t}\right)\mathbf{v}_{m,k}(n)=\frac{w_k(n)\mathbf{G}_{m,k}(n)\mathbf{v}_{m,k}(n)}{1+\mathbf{v}_{m,k}^H(n)\mathbf{G}_{m,k}(n)\mathbf{v}_{m,k}(n)+i_{m,k}(n)}, \quad \square$$

$$k\in B_m(n), \; m=1,\ldots,M, \; n=1,\ldots,N.$$

**Acknowledgements.** The authors would like to acknowledge the supports of Cyber Space Research Institute (CSRI) and I. R. Iran Ministry of Science, Research and Technology.